\documentclass[10pt,twocolumn,prl,aps,floatfix,superscriptaddress,longbibliography]{revtex4-2}

\usepackage{xcolor}
\usepackage{hyperref}

\usepackage{graphicx}
\usepackage{physics}

\usepackage{amsthm}
\usepackage{amsmath}
\usepackage{amssymb}
\usepackage{calligra}

\renewcommand{\vec}[1]{\boldsymbol{#1}}

\usepackage{calligra}
\DeclareMathAlphabet{\mathcalligra}{T1}{calligra}{m}{n}
\DeclareFontShape{T1}{calligra}{m}{n}{<->s*[2.2]callig15}{}

\DeclareSymbolFont{usualmathcal}{OMS}{cmsy}{m}{n}
\DeclareSymbolFontAlphabet{\mathcal}{usualmathcal}

\usepackage{tikz} 
\usetikzlibrary{arrows,scopes} 
\newcommand{\rc}{%
\resizebox{!}{1.25ex}{%
    \begin{tikzpicture}[>=round cap]
        \clip (0.09em,-0.05ex) rectangle (0.61em,0.81ex);
        \draw [line width=.11ex, <->, rounded corners=0.13ex] (0.1em,0.1ex) .. controls (0.24em,0.4ex) .. (0.35em,0.8ex) .. controls (0.29em,0.725ex) .. (0.25em,0.6ex) .. controls (0.7em,0.8ex) and (0.08em,-0.4ex) .. (0.55em,0.25ex);
    \end{tikzpicture}%
}%
}

\usepackage[color=orange!60,textsize=scriptsize]{todonotes}
\usepackage{siunitx}
\DeclareSIUnit\angstrom{\text{Å}}

\usepackage{pdfpages} 
\usepackage{pgffor} 
\usepackage{xr} 

\makeatletter
\AtBeginDocument{\let\LS@rot\@undefined}
\makeatother

\def\supplementfilename{massflow1d_supplement}

\newif\ifarXiv
\arXivtrue 

\ifarXiv
    \pdfximage{\supplementfilename.pdf}
    \def\numbersupplementpages{\the\pdflastximagepages}
\fi

\begin{document}

\title{Friedel oscillations in one-dimensional $^4$He}

\author{Bernd Rosenow}
\affiliation{Institut f\"ur Theoretische Physik, Universit\"at Leipzig, D-04103, Leipzig, Germany}

\author{Adrian Del Maestro}
\affiliation{Department of Physics and Astronomy, University of Tennessee, Knoxville, TN 37996, USA}
\affiliation{Min H.~Kao Department of Electrical Engineering and Computer Science, University of Tennessee, Knoxville, TN 37996, USA}

\begin{abstract}
One-dimensional bosonic systems, such as helium confined to nanopores, exhibit Luttinger liquid behavior characterized by density waves as collective excitations. We investigate the impact of a scattering potential on a low dimensional quantum liquid.  We consider a microscopic model of $^4$He inside a perturbed nanopore with a localized constriction, and employ quantum Monte Carlo simulations to analyze the density of the core within an effective low-energy framework. Our results reveal the emergence of Friedel 
 oscillations in a bosonic quantum liquid without a Fermi surface. Furthermore, we utilize the Luttinger liquid model to predict experimentally observable signatures of this pinning phenomena in elastic scattering and via the temperature and pressure dependence of mass transport through the deformed nanopore.
\end{abstract}
\maketitle

Due to a hardcore constraint, bosonic atoms confined to one dimension (1D) cannot pass each other spatially, and thus can only move collectively. This behavior is described by the paradigm of a Luttinger liquid (LL) characterized by density waves as collective excitations \cite{Haldane:1981gd,Tomonaga1950,Luttinger1963,Giamarchi2003}.  A LL is dominated by strong quantum fluctuations, suppressing most types of order and admitting a description in terms of instabilities.  Due to the charge density wave instability of a LL,  an impurity potential can be amplified \cite{Matveev:1993}, causing long-ranged static density oscillations in response, which are known as Friedel oscillations \cite{Friedel:1958}. In fermionic systems,  such oscillations already occur in the absence of interactions, due to the combination of a Fermi surface and the phase shifts of single-particle eigenstates caused by the impurity potential. For electrons, Friedel oscillations were first observed via tunneling spectroscopy \cite{Hasegawa:1993, Crommie:1993} and are now used as a powerful tool to map out complex Fermi surfaces \cite{Petersen98}. Since bosons do not have a Fermi surface in general, strong interactions in the 1D limit are required, and the potential of using Friedel oscillations to probe the quantum nature of correlated bosonic liquids remains unrealized. 

\begin{figure}[t]
\begin{center}
    \includegraphics[width=\columnwidth]{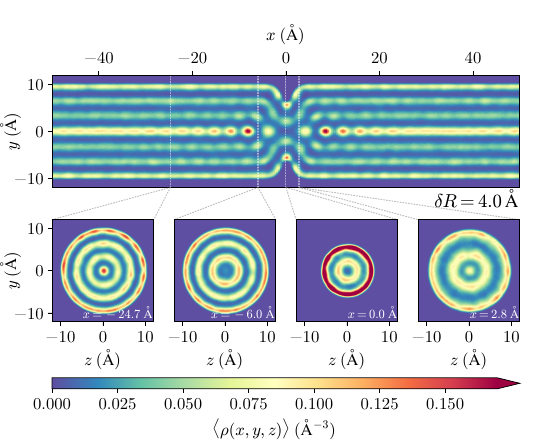}
\end{center}
\caption{Particle density inside the constricted nanopore. The top panel shows a spatial cut of local density $\expval{\rho(x,y,z=0)}$ of $\sim 1000$ $^4$He atoms for a perturbation with $\delta R = \SI{4.0}{\angstrom}$ and $w=\SI{3}{\angstrom}$ at $T=\SI{2}{\kelvin}$.  The call outs show density cross-sections at fixed $x$.  Far from the constriction, we observe a shell structure with a 1D core.  Near the pinch, the screened impurity leads to density (Friedel) oscillations in the core.  \label{fig:densityCutsSlices} 
}
\end{figure}
The quantum liquids of helium have received considerable attention as candidates to observe LL behavior \cite{Savard:2009fc, Savard:2011hs, Duc:2015sa, Velasco:2012de, Velasco:2014fe, Botimer:2016pd, Vekhov:2012mf, Cheng:2015yh, Shin:2017ji, Yager:2013cva, Toda:2007cv, Eggel:2011fj, Taniguchi:2013us, Demura:2015hq, Demura:2017zj, DelMaestro22}, made possible by the ability to fabricate porous networks \cite{Kresge:1992yv} or individual pores \cite{Savard:2009fc} with nanometer radii. The physics of flow through such pores is distinct from superflow through solid helium \cite{Vekhov:2012mf,Cheng:2016cd,Shin:2017ji} believed to be characterized by a transverse quantum liquid \cite{Kuklov:2022st,Radzihovsky:2023se}. Helium confined in nanopores offers an ideal system due to strong interactions, large quantum effects, high purity and remarkable degree of tunability through density, temperature and even the ability to alter particle statistics via the isotopes $^3$He and $^4$He.  In the latter case, numerical simulations have observed a power law decay of correlation functions controlled by the Luttinger parameter $K \simeq 1.3$ for $^4$He confined inside nanopores \cite{DelMaestro:2011ll,Kulchytskyy:2013dh,Bertaina:2016gu,Markic:2018bw}.
There is tantalizing  evidence that the 1D regime is within experimental reach, including the observation of suppressed superfluid critical velocity in single nanopores \cite{Duc:2015sa}, enhanced superfluid dissipation \cite{Toda:2007cv} and an edge singularity in the dynamical structure factor of helium confined to porous media \cite{DelMaestro22}.  

Even in carefully synthesized or microfabricated porous systems, nanopores are not perfect cylinders and instead contain modulations in their radius \cite{Sonwane:1999sc,Kim:2007}.  It is natural to explore if such constrictions in an interacting bosonic quantum liquid may lead to pinning and backscattering. In this study we use quantum Monte Carlo (QMC) simulations
to  identify Friedel oscillations in an atomic Luttinger liquid of $^4$He in such a non-ideal pore, characterized by a local constriction.  By comparing QMC observations of density oscillations with predictions for the  LL model, we determine the effective impurity strength in the latter. We find that even strong constrictions are screened by the $^4$He shell structure inside the pore, leading to the 1D core being in a weak pinning regime (see Figure~\ref{fig:densityCutsSlices}).   
Our LL characterization of Friedel oscillations leads yields experimentally testable predictions for elastic scattering  and for the temperature and pressure dependence of mass transport through nanopores.  Confirmation would provide conclusive experimental proof of linear quantum hydrodynamics for confined $^4$He.

\emph{Microscopic Model} -- We consider a system of $N$ $^4$He atoms confined inside cylindrical nanopores of length $L$ (aligned along the $x$-direction with periodic boundary conditions) subject to a constriction localized at $x=0$ described by the Hamiltonian:
\begin{equation}
    H = -\frac{\hbar^2}{2m}\sum_{i=1}^N \nabla_i^2 + \sum_{i=1}^N U_{\rm
    pore}(R,\vec{r}_i) + \frac{1}{2}\sum_{i,j} V(\vec{r_i}-\vec{r_j}).
\label{eq:Ham}
\end{equation}
Atoms of mass $m$ and positions $\vec{r}_i = (x_i,y_i,z_i)$
interact through $V$ which is known to high precision \cite{Aziz:1979hs,
Przybytek:2010js, Cencek:2012iz} and with the walls of the nanpore through $U_{\rm pore}$. Unlike previous simulations inside perfect cylindrical pores  of radius $R_0$ \cite{Chakravarty:1997qa,Gordillo:2000qo, Boninsegni:2001ho, Rossi:2005ll,
DelMaestro:2011ll,DelMaestro:2012td, Kulchytskyy:2013dh, Kiriyama:2014pi,
Markic:2015bu, Markic:2018bw, Markic:2020ip, Nichols:2020of, Nava:2022qo},
here we introduce a smooth constrictive deformation of the radius of magnitude $\delta R$ and width $w$ which gives the pore an \emph{hourglass} shape defined by an $x$-dependent radius:
\begin{equation}
    R(R_0,\delta R,w;x) = R_0 + \delta R\qty[\frac{\tanh^2\qty({2x}/{w})}{\tanh^2(L/w)}-1].
    \label{eq:Rx}
\end{equation}
This choice is motivated by electron microscopy on fabricated nanopores \cite{Kim:2007}. For $\delta R/R_0 \ll
1$ and $w/L \ll 1$, the hourglass confinement potential $U_{\rm pore}$ is shown in Fig.~\ref{fig:hourglass} with details given in the supplement \cite{supplemental}.
\begin{figure}[t]
\begin{center}
\includegraphics[width=\columnwidth]{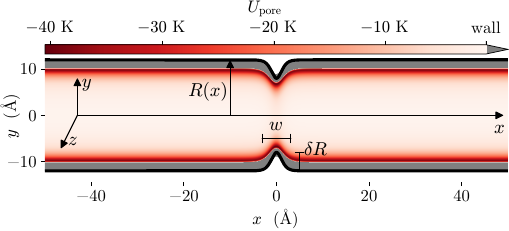}
\end{center}
\vspace{-1.5em}
\caption{Hourglass geometry and potential.  A cross-section of the axially symmetric hourglass potential in the $z=0$ plane for a nanopore of length $L=\SI{100}{\angstrom}$ and radius $R_0 = \SI{12}{\angstrom}$ with a constriction defined by Eq.~(\ref{eq:Rx}) with $\delta R = \SI{4}{\angstrom}$ and $w = \SI{3}{\angstrom}$.  While the black line denotes the actual radius function $R(x)$, the grey shaded region in the colorbar represents the hard-wall of the hourglass potential where $U_{\rm pore} \gg \SI{1}{\kelvin}$. 
\label{fig:hourglass} 
}
\end{figure}

For cylindrical pores ($\delta R=0$), the interplay between confinement and interparticle interactions produces a set of nested cylindrical shells with the outer ones remaining solid due to their proximity to the wall \cite{DelMaestro:2011ll}. At low enough temperature ($T \lesssim \SI{2}{\kelvin}$) the innermost shells can become superfluid \cite{Kulchytskyy:2013dh}, and for fine-tuned radii, a central core of atoms can exist describable by the gapless linear quantum hydrodynamics of LL theory with Luttinger parameter of $K=1.3\pm 0.1$ \cite{DelMaestro:2011ll}.  This is close to that of hard-core bosons ($K = 1$) consistent with what is found for strictly 1D helium chains \cite{Bertaina:2016gu}. In the presence of a constriction, these atoms may be subject to an effective localized potential resulting in the possibility of pinning and backscattering in the LL.   

{\em Numerical Results} -- 
We simulate the Hamiltonian Eq.~\eqref{eq:Ham} via a continuous space worm algorithm path integral quantum Monte Carlo (QMC) method \cite{Boninsegni:2006ed,pimcrepo} measuring the local density of particles $\expval{\rho(\vec{r})} = \expval{ \sum_{i=1}^N \delta (\vec{r}-\vec{r}_i)}$ inside the perturbed nanopore. Here, $\expval{\dots}$ represents a grand canonical thermal expectation value at fixed chemical potential $\mu$ and temperature $T$ as well as an average over potential disorder due to different arrangements of adsorbed atoms in the vicinity of the constriction. For complete details of our simulations and analysis see Ref.~\cite{supplemental}.  Figure~\ref{fig:densityCutsSlices} shows 
the density in the  $z=0$ plane for  a $R_0 = \SI{12.0}{\angstrom}$ nanopore with a deformation $\delta R = \SI{4.0}{\angstrom}$ and width  $w=\SI{3}{\angstrom}$ constriction (chosen to correspond to the approximate van der Waals radius of $^4$He).  
At distances along the pore that are far from the constriction there is a well defined quasi-1D dimensional core of atoms and surrounding shells as in the aforementioned $\delta R=0$ case. Near the perturbation these 
shells bend inward, causing a  drastic density modulation which is in phase with oscillations farther from the constriction.

\begin{figure}[t]
\begin{center}
    \includegraphics[width=\columnwidth]{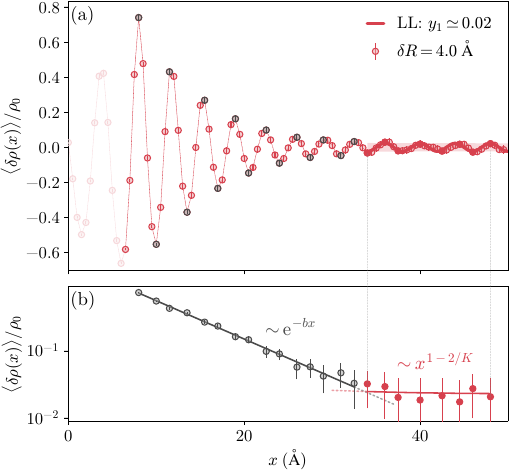}
\end{center}
\caption{
Friedel Oscillations in Nanopores. The dimensionless 1D atomic density within the core $\delta\rho(x)/\rho_0 = \rho(x)/\rho_0-1$ for a constriction with $\delta R = \SI{4.0}{\angstrom}$ at $T=\SI{2.0}{\kelvin}$. (a) Quantum Monte Carlo data (points) with density oscillations near the constriction (faint) are initially suppressed due to the merging of surrounding cylindrical shells, but rebound for $x > w$.  Outside the core, two distinct regions are observed. A rapid decay settles into oscillations with nearly constant amplitude for $x > \SI{33}{\angstrom}$ (shaded). The dashed line is a guide to the eye. (b) The two regimes can be seen more clearly when plotting the oscillation envelope (grey and solid red points from (a)) on a semi-log scale. The solid red line always represents the Luttinger liquid prediction for Friedel oscillations from Eq.~\eqref{friedel.eq} using $y_1$ extracted from simulations and the known value of $K \simeq 1.3$.  
\label{fig:friedel} 
}
\end{figure}

Density oscillations in the core can be quantitatively analyzed by defining a cutoff radius $R_c = \SI{1.75}{\angstrom}$ corresponding to the mid-point position between the core and first surrounding cell where the particle density vanishes for $x \gg w$ as can be seen in the bottom-left panel of Fig.~\ref{fig:densityCutsSlices}.  The resulting 1D density of the core is
\begin{equation}
    {\rho(x)} = \int_0^{2\pi} \! \! \! \! d\varphi \int_0^{R_c} \!\!  \! \! \rc \dd{\rc} {\rho(x,\rc,\varphi)}
    \label{eq:rho1D}
\end{equation}
using cylindrical coordinates where $\rc = \sqrt{y^2 + z^2}$ and $\varphi = \arctan(y/z)$. QMC results for Eq.~\eqref{eq:rho1D} are shown in Fig.~\ref{fig:friedel} where $\delta\rho(x) \equiv \rho(x) - \rho_0$ with $\rho_0 = L^{-1}\int_{-L/2}^{L/2}\rho(x)$. 
Panel (a) shows the $x$-dependent density oscillations observed in the core, with the full red line (in the highlighted region) demonstrating the LL prediction from Eq.~\eqref{friedel.eq} (see next sections for details). For an unperturbed pore with $\delta R = 0.0$, Galilean invariance restricts $\rho(x) = \rho_0$ \cite{DelMaestro:2011ll} and no oscillations are expected.  The envelope of oscillations on a semi-log scale shows two distinct decay regimes in panel (b), with the second arrested decay (due to the presence of periodic boundary conditions and small exponent) admitting a quantitative analysis within LL theory.

\emph{Luttinger Liquid Model} --
The low energy physics of 1D $^4$He is described by a LL with Hamiltonian  
\begin{align}
    \mathcal{H} &= \frac{v}{2\pi} \int_{-L/2}^{L/2}\!\!\! d x \left[ \frac{1}{K}
    (\partial_x \phi)^2 + K (\partial_x \theta)^2 \right] + \mathcal{H}_{ps} 
    \label{eq:H} 
    \end{align}
    with $\partial_x \theta/\pi$ being conjugate to $\phi$ such that 
$[\partial_x \theta(x),\phi(x')] = i\pi\delta(x-x')$.  We use units in which $\hbar=1$ and $k_{\rm B} = 1$ and $v$ is the phonon velocity.
The field $\phi(x)$ describes the quantum phase, and the particle density can be expressed in terms of the field $\theta(x)$ as
%
\begin{equation}
\rho(x) \simeq \left[ \rho_0 + {1 \over \pi} \partial_x \theta(x)\right] \qty[1 + 2\cos2\qty(\theta(x) + \pi \rho_0 x)] 
\label{particledenstiy.eq}
\end{equation}
%
where oscillatory terms with higher multiples of $\pi \rho_0$ are neglected. 
The interaction of helium atoms with an external potential is modeled by $\mathcal{H}_{ps} = \int_{-L/2}^{L/2} dx \rho(x) \mathcal{V}(x)$ where we consider a symmetric potential $\mathcal{V}(x)=\mathcal{V}(-x)$ centered at $x=0$. In the microscopic model of helium atoms, the external potential gives rise to  backscattering of particles, whereas in the LL model it causes phase slips of $2 \pi$ in the phase. Since the density operator Eq.~(\ref{particledenstiy.eq}) contains oscillatory contributions, it is useful to characterize the scattering potential in terms of its most dominant Fourier coefficient
%
\begin{equation}
    y_1 = { 1 \over v} \int_{-L/2}^{L/2} dx\, \mathcal{V}(x)\, \cos(2 \pi \rho_0 x) \ \ ,
    \label{eq:y1}
\end{equation}
%
which is called a fugacity in LL parlance.  The phase slip Hamiltonian can be expressed as
    %
$
    \mathcal{H}_{ps}  =   \frac{2 v}{a} y_1 \cos \left[2\theta(0) \right] \ ,
    $
    where $a =1/\rho_0$ is the average inter-particle distance. This expression can be used in connection with QMC results displayed in Fig.~\ref{fig:friedel} to extract the fugacity $y_1$.

{\em Density Oscillations} --
For a perfect LL, a localized potential induces Friedel oscillations of the particle density. The leading contribution of these density oscillations is due to the $2 k_F$-component $y_1$ of the potential and has been derived in Ref.~\cite{Egger95} as  %
\begin{equation}
    \frac{\expval{\delta \rho(x)}}{\rho_0} = - {2^{2\over K} K y_1 \over 2 \pi} B\left({1 \over 2},{1 \over K} - {1 \over 2}\right) {\cos\left(2 \pi \rho_0 x \right) \over 
\left(1 + {x \over \alpha}\right)^{{2 \over K}-1}}
\label{friedel.eq}
\end{equation}
where $\alpha = a K/2 \pi$ is a renormalized microscopic length scale and $B(\cdot,\cdot)$ is the Beta function \cite{abramowitz1965handbook}. In the presence of periodic boundary conditions, decays are governed by a chord length $x \to L/\pi \sin(\pi x/L)$. Eq.~\ref{friedel.eq} is valid for a weak pinning potential, and more specifically in the regime $x \ll x_0$, with $x_0 = (aK^2/4\pi^2)(4y_1)^{K/(1-K)}$ \cite{Egger1996}.  

Using the QMC data shown in Fig.~\ref{fig:friedel}, we directly extract $y_1 \simeq 0.02$ from the envelope of density oscillations in the weakly decaying regime 
using the previously known value of $K\simeq 1.3$ \cite{DelMaestro:2011ll} for an unperturbed nanopore of the same radius.  This value yields a crossover scale $x_0 \approx 2400 a$, much larger than the scale of our simulations.  This confirms that the pinning effect is weak, supporting the use of the perturbative formula to describe the envelope of the density oscillations.  It is important to note that the thermal length $\ell = \pi v/T \approx 15 a$ -- calculated using $v = 42 \si{\angstrom\kelvin}\approx \SI{550}{\meter \per \second}$ \cite{DelMaestro:2012td} --   does not significantly influence the observed oscillations.  This suggests that at the studied scale, thermal fluctuations are not the dominant factor affecting the decay of the oscillations, allowing observation of quantum effects. 

Having theoretically demonstrated the existence of Friedel oscillations in a interacting system of bosons, we now discuss two experimental signatures of this effect that should be observable in confined $^4$He - a unique divergence observable in the static structure factor and the temperature dependence of mass flow through nanopores.

{\em Signatures in Elastic Scattering} --
Static charge order induced by Friedel oscillations pinned to the positions of constrictions contributes to the dynamic structure factor at zero frequency, $S(q,0)$.  Denoting the Fourier transform of the charge distribution due to a single constriction by $\delta \rho (q)$, the contribution is
%
\begin{equation}
    S_{c} (q,0)  \ = \ N_{c} \, \left|\langle \delta \rho (q)\rangle \right|^2 \ \ .
    \label{eq:Scq}
\end{equation}
%
Here, $N_c$ denotes the total number of constrictions,   where the signal arises due to the ensemble of impurities present in the sample, equivalent to the average over quenched disorder present in our numerical simulations. When evaluating Eq.~\eqref{eq:Scq} using the density oscillations Eq.~(\ref{friedel.eq}), 
one finds (see Ref.~\cite{supplemental}) a divergent peak $S_c(q,0) \propto 1/|q - 2 k_{\rm F}|^{4 - 4/K}$, which is qualitatively different 
from the finite cusp at the same momentum $2 k_{\rm F} = 2\pi\rho_0$  from the connected part of the density-density correlation function \cite{DelMaestro22}.
In a real experiment, this divergence should be observable in elastic scattering of helium atoms confined inside porous media, manifest as a peak whose height grows with $1/T^{4 - 4/K}\simeq 1/T$ for the pores considered here with $K\simeq 1.3$.

\begin{figure}[t]
\begin{center}
\includegraphics[width=\columnwidth]{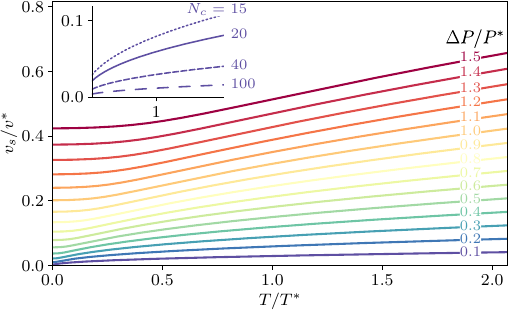}
\end{center}
\caption{
\label{fig:LLTransport} 
 Superfluid flow velocity in a Luttinger liquid with $K=1.3$ and $N_c = 20$ impurities as a function of temperature $T$ for different pressure differences $\Delta P$. All quantities are measured with respect to an experimentally relevant scale (indicated by $\ast$) set by fixing the pressure difference between the bulk reservoirs at both ends of the LL to be 1~atm.  The inset shows the effect of modifying $N_c$ for $\Delta P / P^\ast = 0.1$.
}
\end{figure}

{\em  Signatures in Transport} --
Friedel oscillations will have an effect on the temperature dependence of the superfluid mass flow $\rho_0 v_s$ (where $v_s$ is the superfluid velocity) for helium confined at the core of deformed nanopores subject to a pressure difference $\Delta P$.  A strength of our LL analysis, is the ability to make non-equilibrium predictions in this regime which are not accessible in numerical simulations. 
The pressure difference is related to the phase slip rate via 
$\Delta P / \rho_s =  \Delta \dot{\phi}$, using the Gibbs-Duhem relation with $\rho_s$  the 3D superfluid density and the superfluid current is defined in terms of the phase $\phi$ of the condensate via $j = v\partial_x \phi / {\pi K}$.
Integrating the expression for the current over the length of the pore we obtain the phase difference, and taking the expectation value of its time derivative, the average phase slip rate is
\begin{align}
    \Delta \dot{\phi} = \frac{ \pi K }{ v} \int_{-L/2}^{L/2}\!\!\! d x
    \left \langle \frac{d}{d t} j(x,t) \right \rangle_{v_s} \ .
    \label{eq:DeltaPhi}
\end{align}
Here, $\langle \cdots \rangle_{v_s}$
indicates that the expectation value is taken with respect to a background
superfluid flow defined by the boundary condition
   $ \langle j(x=0,t) \rangle =  \rho_0 v_s$.
Evaluating Eq.~\eqref{eq:DeltaPhi} leads to (details in the supplement \cite{supplemental}):
\begin{equation}
\Delta\dot{\phi} =  y_1^2 N_c \rho_0 v_s \left(\frac{a T}{v}\right)^{2/K - 2} 
\Phi \left(\frac{ \rho_0 v_s}{T}\right) \ ,
\label{eq:PhiDotScaling}
\end{equation}
where we assume that the length of the pore is much longer than any
thermal length: $L \gg \ell$ in the temperature range of interest, and the $N_c$ independent constrictions are spaced more than the thermal length apart. The scaling function $\Phi$ has the limiting behavior 
$\Phi(z\to0) = {\rm const.}$  and $\Phi(z\to\infty) \propto z^{2/K-2}$. Eq.~\eqref{eq:PhiDotScaling} provides the non-linear pressure-velocity relation analogous to a current-voltage characteristic in a wire.  To extract the temperature dependence from Eq.~(\ref{eq:PhiDotScaling}), it is useful to introduce temperature and velocity scales set by the external pressure difference: we set a pressure scale $P^*= 1~{\rm atm}$ leading to $T^\ast \simeq \SI{0.3}{\kelvin}$ and $v^\ast \simeq \SI{17}{m/s}$.  The resulting temperature dependence is shown in Fig.~\ref{fig:LLTransport} for different values of the 
pressure difference $\Delta P$.  
At low driving pressure (or for many constrictions), we predict enhanced scattering (non-linear suppression of massflow) at low temperatures on the order of a tens of \si{\milli\kelvin} due to the formation of Friedel oscillations, consistent with the temperatures needed to observe quantum phase slips in bulk superfluid helium \cite{Varoquaux:2015}. This is qualitatively different from the enhancement of superflow below $T_\lambda$ in wider constrictions \cite{Vekhov:2012mf, Duc:2015sa, DelMaestro:2017kx} in which dissipation is induced by thermal phase slips. To understand the behavior for high driving pressures, we work in a regime where the argument of the scaling function is large,  and the relation between superfluid velocity and phase slip rate is non-linear:
%
\begin{equation}
 \left({v_s\over v}\right)^{{2 \over K} -1}  = \ {\Delta \dot{\phi}\over \rho_0 v} \left[ y_1^2 N_c \left( 2\pi\right)^{2 \over K}  \sin(\tfrac{2\pi}{K})
\Gamma (1 - \tfrac{2}{K})\right]^{-1} \, .
\end{equation}
%
The observation of strong pressure dependence ($v_s \propto \Delta P^2$ for $K \simeq 1.3)$ of $v_s$ would indicate that an experiment has entered a quantum dissipative regime, distinct from the logarithmic pressure dependence arising from thermally activated phase slips \cite{Duc:2015sa,DelMaestro:2017kx}.

{\em Discussion} --
In this study, we explored $^4$He confined inside hourglass-shaped nanopores to model the disorder relevant in experiments on low-dimensional superfluids.  By analyzing large scale quantum Monte Carlo simulations within the context of a Luttinger liquid (LL) subject to an impurity,  we parametrize the effective theory and provide a microscopic view of the induced Friedel oscillations in a bosonic liquid. This confirms the robustness of the LL framework in describing confined quantum fluids. 

Our predictions have implications for experiments. The existence of Friedel oscillations in the density contributes an additional signal to the elastic scattering with a power-law divergence,  which could be detected in helium confined inside nanoporous media. Recent neutron scattering experiments in such a system \cite{DelMaestro22} found increased scattering at $2k_{\rm F}$ both in the elastic and inelastic channel. A similar setup could probe the predicted divergence of the elastic signal.  

Using the extracted backscattering amplitude for a constriction, we find that both thermally and pressure limited superflow through nanopores is within experimental reach,  including transport in single pores \cite{Duc:2015sa,Botimer:2016pd}, or oriented nanoporous media  \cite{Kaneko:2023sm}.  In the quantum regime, the phase slip fugacity is a relevant perturbation, which localizes the superfluid at temperatures below $\SI{100}{\milli\kelvin}$ offering an unambiguous signal of Luttinger liquid behavior in confined helium.

\emph{Code Availability} -- All software and data used in this study are available online \cite{paperrepo}.

\acknowledgments

We thank P.~Sokol, P.~Taborek, G.~Gervais, and J.~Demidio for useful discussions. A.D. acknowledges support from the U.S. Department of Energy, Office of Science, Office of Basic Energy Sciences, under Award Number DE-SC0024333.

\bibliography{refs_massflow1d}

\ifarXiv


\section*{Supplementary material}

# Supplemental Material for: "Friedel oscillations in one-dimensional $^4$He"

Bernd Rosenow and Adrian Del Maestro

## I. CONFINEMENT POTENTIAL INSIDE HOURGLASS SHAPED NANOPORES

To compute the confinement potential $U_{\rm pore}$ in Eq. (1) in the main text we assume an infinitely long cavity composed of cylindrical slices of different radii $R(x)$ carved inside a uniform medium [S1]:

$$U_{\rm pore}(R;\boldsymbol{r}) = \frac{\pi n \varepsilon \sigma^3}{3}\left[\left(\frac{\sigma}{R(x)}\right)^9 u_9\left(\frac{\sqrt{y^2+z^2}}{R(x)}\right) - \left(\frac{\sigma}{R(x)}\right)^3 u_3\left(\frac{\sqrt{y^2+z^2}}{R(x)}\right)\right] \quad (S1)$$

with

$$u_3(\varkappa) = \frac{2}{(1-\varkappa^2)^3}\left[(7+\varkappa^2)E(\varkappa) - 4(1-\varkappa^2)K(\varkappa)\right] \quad (S2)$$

$$u_9(\varkappa) = \frac{(1091 + 11156\varkappa^2 + 16434\varkappa^4 + 4052\varkappa^6 + 35\varkappa^8)E(\varkappa) - 8(1-\varkappa^2)(1+7\varkappa^2)(97+134\varkappa^2+25\varkappa^4)K(\varkappa)}{240(1-\varkappa^2)^9} \quad (S3)$$

where $\boldsymbol{r} = (x,y,z)$, $\varkappa = \sqrt{y^2+z^2}$ is the distance of the atom from the pore axis, $n$ is the density of the confining media, $\varepsilon$ is the strength of the interaction, $\sigma$ is the hard core distance, and $K(r)$ and $E(r)$ are complete elliptic integrals of the first and second kind respectively. Using the Lorentz-Berthelot mixing rules for $^4$He interacting with amorphous silicate with density $n = 0.078\,\text{Å}^{-3}$ [S2–S4] we set $\varepsilon = 10.22\,\text{K}$ and $\sigma = 2.628\,\text{Å}$. A heatmap of the hourglass potential for $w = 3.0\,\text{Å}$ and $\delta R = 4.0\,\text{Å}$ is shown in Fig. 2 in the main text, while Fig. S1 shows some cuts at different positions $x$ along the pore.

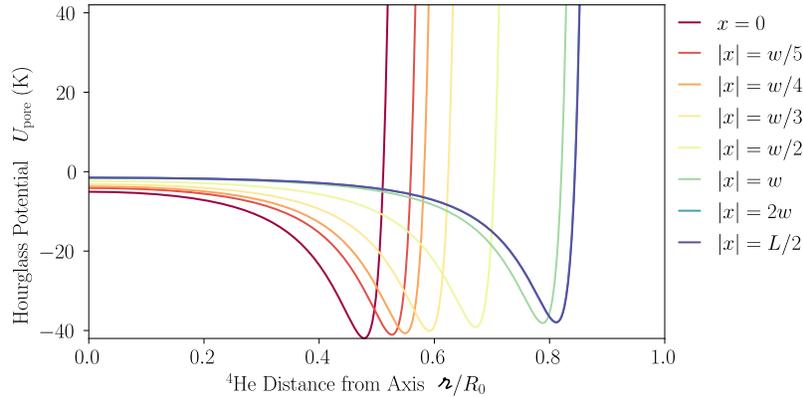

FIG. S1. The hourglass potential $U_{\rm pore}$ for different positions along the pore as a function of the distance from the axis for $\delta R = 4.0\,\text{Å}$ and $w = 3.0\,\text{Å}$. Effects are strongest near $x=0$, showing a large shift of the potential minimum (when compared to the smooth pore when $|x| \gg w$) and the addition of a constant shift near the pore center which acts like an additional chemical potential.

Complete microscopic details along with software needed to compute the potential can be found in a github repository [S5].

## II. PATH INTEGRAL QUANTUM MONTE CARLO

### A. Simulation Details

A system of $N$ confined helium described by Eq. (1) in the main text was simulated using a quantum Monte Carlo algorithm exploiting path integrals [S6–S8] utilizing our open source software [S9].

| Parameter | Symbol | Value |
|---|---|---|
| Temperature | $T$ | $2.0\,\mathrm{K}$ |
| Chemical Potential | $\mu$ | $-7.2\,\mathrm{K}$ |
| Imag. Time Step | $\tau$ | $0.004\,\mathrm{K}^{-1}$ |
| Pore Length | $L$ | $100\,\mathrm{\AA}$ |
| Pore Radius | $R_0$ | $12\,\mathrm{\AA}$ |
| Hourglass Pinch Size | $\delta R$ | $4.0\,\mathrm{\AA}$ |
| Hourglass Pinch Width | $w$ | $3.0\,\mathrm{\AA}$ |
| Number Seeds | $\mathcal{N}$ | 150 |
| Number Measurements per Seed | $M$ | 100,000 |

TABLE S1. Path integral quantum Monte Carlo simulation parameters.

Grand canonical finite temperature expectation values of any observable $\hat{\mathcal{O}}$ inside the hourglass shaped nanopore were computed using

$$\left\langle \hat{\mathcal{O}} \right\rangle = \frac{1}{\mathcal{Z}} \operatorname{Tr}\left\{ \hat{\mathcal{O}}\, \mathrm{e}^{-\beta(H-\mu N)} \right\} \tag{S4}$$

where $\beta = 1/T$ is the inverse temperature (in units where $k_\mathrm{B} = 1$), $\mu$ is the chemical potential, $N$ is the particle number operator, and the partition function $\mathcal{Z} = \operatorname{Tr} \mathrm{e}^{-\beta(H-\mu N)}$ can be evaluated by Monte Carlo sampling discrete imaginary time paths (worldlines) over the set of all permutations of the indistinguishable $^4$He atoms. Due to the bosonic nature of helium-4, expectation values are stochastically exact up to timestep errors introduced via an $\mathrm{O}(\tau^4)$ approximation for the short imaginary time propagator $\mathrm{e}^{-\tau H}$ [S10, S11]. We choose a value of $\tau$ such that any Trotter errors are smaller than statistical uncertainty in averages. All simulation parameters are detailed in Table S1 and we consider a perturbed hourglass shaped cylinder with periodic boundary conditions in the $x$-direction. Further details (including the effects of the length of a smooth cylindrical pore) are reported in Refs. [S8, S12], where emergent quantum hydrodynamics for core atoms were identified in systems where the aspect ratio of radius:length is greater than 8:1 as is the case here.

### B. Density Averaging

As we are interested in the emergence of static density oscillations as a result of the nanopore constriction, we measure the average spatially resolved density:

$$\langle \rho(\boldsymbol{r}) \rangle = \left\langle \sum_{i=1}^{N} \delta(\boldsymbol{r} - \boldsymbol{r_i}) \right\rangle. \tag{S5}$$

Due to the grand canonical nature of the simulations (with fixed chemical potential $\mu$), we find $\langle N \rangle = \int \mathrm{d}\boldsymbol{r}\, \langle \rho(\boldsymbol{r}) \rangle \approx 1000$ inside the perturbed nanopore. The 1D density of the core can be examined by integrating Eq. (S5) over a region of the hourglass as described by Eq. (3) in the main text.

There are two features of our simulations that warrant further comment, both related to the interplay between the externally imposed geometry of the nanopore set by $L, R_0, \delta R$ and $w$, and the intrinsic length scale $r_\mathrm{He}$ which corresponds to the minimum of the helium-helium interaction $V$ in Eq. (1). First: as seen in Fig. S1, the minimum of the adsorption potential $\min_{r} U_\mathrm{pore} \approx -40\,\mathrm{K}$ has a magnitude much larger than temperature. Thus, after an initial burn-in period where the system equilibrates by gradually increasing the number of atoms in the pore (physically analogous to a capillary effect), a series of concentric shells [S8] are formed and atoms near the wall *freeze* into a perturbed quasi-2D cylindrical shell mirroring the hourglass shape. This outer shell is mostly static over the Monte Carlo timescale of our simulations, contributing a complicated many-body effective potential experienced by atoms in the core akin to a quenched contact potential localized near $x = 0$. As a result, we perform $\mathcal{N}$ independent simulations starting from different initial configurations (seeds), and report averages over seeds along with the standard error in the mean. While this is standard practice to improve statistics, there is no guarantee that atoms in the center of the

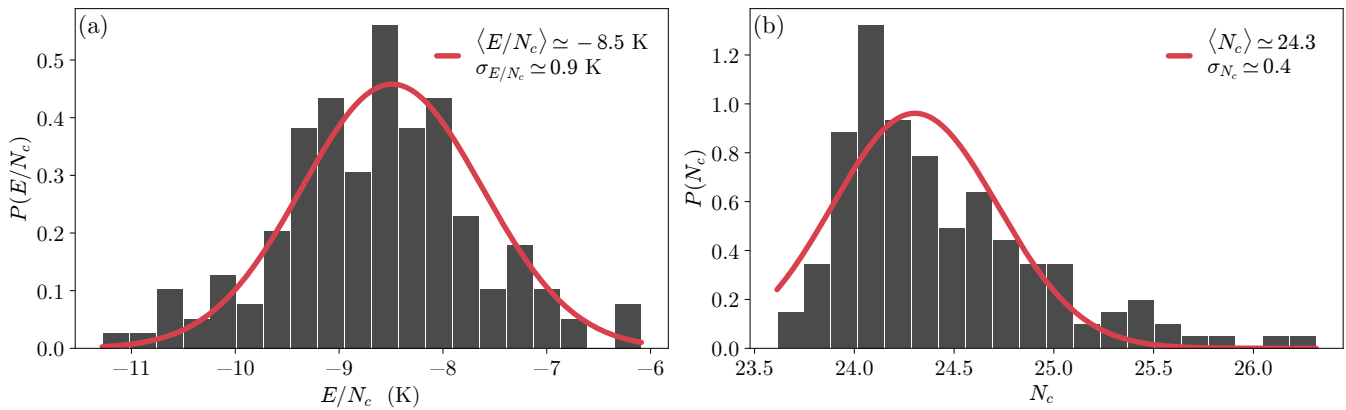

FIG. S2. The full distribution of the energy per particle (a) and total number of particles (b) over 150 independent seeds. The solid lines are best-fits to a normal distribution with mean and standard deviation indicated in the legend.

hourglass are governed by the same Boltzmann distribution and this can be confirmed by examining histograms of the average energy per particle or the average number of particles in the core over seeds as seen in Fig. S2. While the energy per particle is reasonably well described by a normal distribution, the existence of fat tails towards large $N_c$ in panel (b) brings us to the second comment. In the grand canonical ensemble, a translationally invariant 1D chain of helium at a chemical potential corresponding to saturated vapor pressure will exhibit quantum liquid behavior with an average density $\langle \rho(x) \rangle \approx 1/r_{\rm He} \approx 0.3\,\text{Å}^{-1}$ [S12]. Thus, for a nanopore with fixed length $L$ and periodic boundary conditions along the axis, in the absence of a fine-tuned length, one can observe fluctuations in the core particle number due to a mismatch between $L$ and $r_{\rm He}$. This effect is accentuated in our system due to the transverse spread of the $^4$He wavefunctions in the core and the effective potential near the constriction which reduces the local density near $x = 0$ as seen in Figure 1 of the main text. For a given finite size simulation (corresponding to a single seed), this may lead to a difficulty of observing a simple signature of static pinned density oscillations within a reasonable computational budget as can be seen in Fig. S3. However, as mentioned above, this difficulty can be overcome by

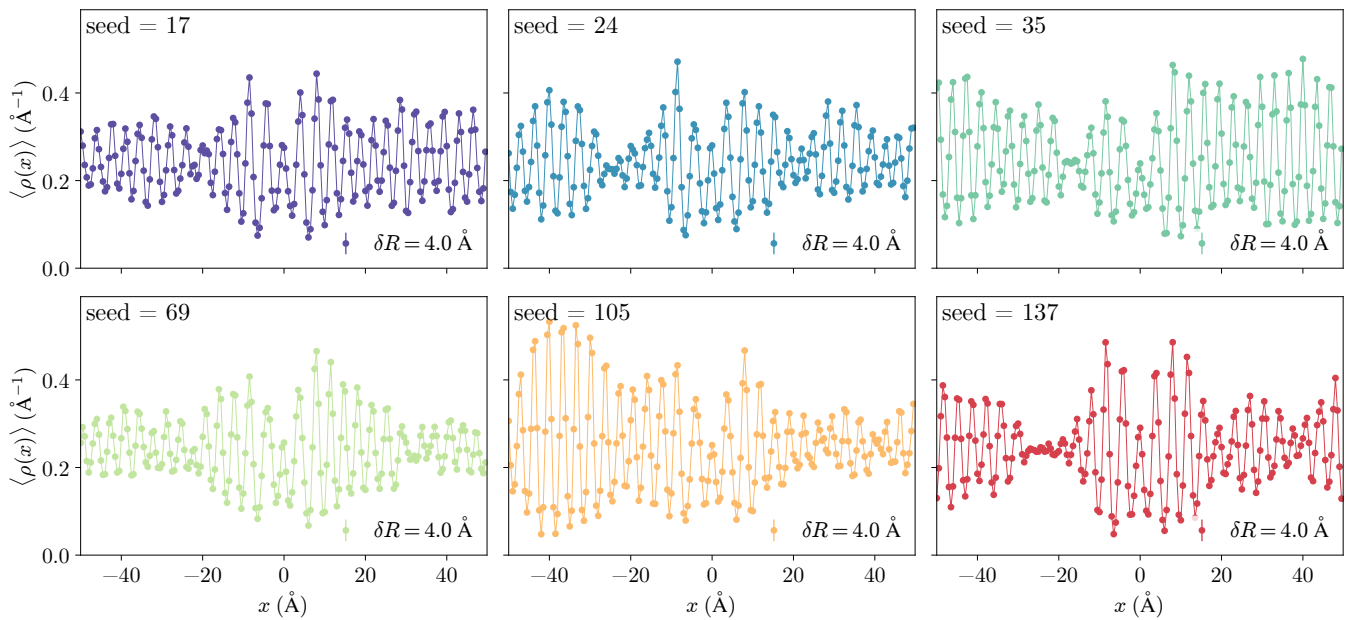

FIG. S3. The average density in the core region along the axis of the hourglass shaped nanopore as measured via Eq. (3) of the main text for 6 (of 150) random seeds. Due to finite quantum Monte Carlo run times, each seed acts as a realization of potential disorder due to the large energy barrier to equilibrate $^4$He atoms near the constriction.

averaging over random seeds (each effectively corresponding to a different realization of quenched potential disorder) yielding the Friedel oscillations observed in Fig. 3 in the main text.

### III.  SCATTERING CALCULATION

The dynamic structure factor at zero frequency is defined as the expectation value

$$S(q, \omega = 0) = \langle \rho(-q, \omega = 0) \rho(q, \omega = 0) \rangle \ . \tag{S6}$$

In the absence of Friedel oscillations due to constrictions, there is only a connected contribution to the above expectation value, which was studied in Ref. [S13]. Here we focus on the case of Friedel oscillations, which contribute a finite $\langle \delta \rho(q, \omega = 0) \rangle$, which can be computed as a Fourier transform of Eq. (7) in the main text. We consider an ensemble of $N_c$ constrictions centered at positions $\tilde{x}_i$, and compute their contribution to the dynamic structure factor. In the following, we drop the explicit argument $\omega = 0$ of densities and just consider spatial Fourier transforms. Then, we find

$$\begin{aligned} \langle \rho(q) \rangle &= \sum_{i=1}^{N_c} \int dx \, \langle \delta\rho(x) \rangle \, e^{-iq(x+\tilde{x}_i)} \\ &= \sum_{i=1}^{N_c} e^{-iq\tilde{x}_i} \langle \delta\rho(q) \rangle \ . \end{aligned} \tag{S7}$$

The disconnected contribution to the dynamic structure factor is then given by

$$\begin{aligned} S_c(q,0) &= \sum_{i,j=1}^{N_c} e^{-iq(\tilde{x}_i - \tilde{x}_j)} \langle \delta\rho(-q) \rangle \langle \delta\rho(q) \rangle \\ &\simeq \sum_{i=1}^{N_c} |\langle \delta\rho(q) \rangle|^2 \\ &= N_c |\langle \delta\rho(q) \rangle|^2 \ . \end{aligned} \tag{S8}$$

The Fourier transform of the density Eq. (7) can be expressed in terms of hypergeometric functions, and one finds that the leading contribution diverges in the vicinity of $2k_F$ as

$$\langle \delta\rho(q) \rangle = -\frac{2^{\frac{2}{K}-2} K y_1}{|q - 2k_F|^{2 - 2/K}} B\left(\frac{1}{2}, \frac{1}{K} - \frac{1}{2}\right) \cos\left(\frac{\pi}{K} + |q - 2k_F|\right) \frac{\csc\left(\frac{2\pi}{K}\right)}{\Gamma\left(-1 + \frac{2}{K}\right)} + O[1] \ . \tag{S9}$$

Taking the modulus square of the above expression yields the divergence $S_c(q,0) \propto 1/|q - 2k_F|^{4 - 4/K}$ quoted in the main text.

### IV.  TRANSPORT CALCULATION

The dynamics of the current $j$ are described by an equation of motion $dj/dt = -i[j, H]/\hbar$:

$$\frac{d}{dt} j(x, t) = \frac{4v^2}{Ka} \delta(x) y_1 \sin[2\theta(x)] \ , \tag{S10}$$

where we specialize to the most relevant backscattering mechanism. We now use a path integral representation of the Keldysh formalism, and implement the aforementioned boundary condition within a saddle point approximation. Then, we take the quantum statistical average of Eq. (S10), and obtain to leading order in the potential

$$\Delta \dot{\phi} = \frac{2\pi v^2 y_1^2}{a^2} \int_0^\infty dt \sin(2\pi \rho_0 v_s t) \chi^R(0, t) \ . \tag{S11}$$

Here, $\chi^R(x,t) = -i\Theta(t)\left\langle\left[e^{2i\theta(x,t)}, e^{-2i\theta(0,0)}\right]\right\rangle$ where $\Theta(t)$ is the Heaviside step function [S14]. For the correlation function $\chi^R(x,t)$ we use [S14]

$$\chi^R(x,t) = \frac{2\left(\frac{\pi a T}{v}\right)^{2/K}\Theta(v^2t^2-x^2)\sin\left(\frac{\pi}{K}\right)}{|\sinh(\pi T(t-x/v))\sinh(\pi T(t+x/v))|^{1/K}}. \tag{S12}$$

Inserting the above expression into Eq. (S11) and performing the time integral, we can express the phase slip rate in the scaling form

$$\Delta\dot\phi = y_1^2 N_c \rho_0 v_s \left(\frac{aT}{v}\right)^{2/K-2}\Phi\left(\frac{\rho_0 v_s}{T}\right). \tag{S13}$$

The scaling function in this expression is given by

$$\Phi(z) = \frac{2(2\pi)^{2/K}}{z}\sin\left(\frac{\pi}{K}\right)\mathrm{Im}\left[\frac{\Gamma\left(\frac{1}{K}-iz\right)}{\Gamma\left(1-\frac{1}{K}-iz\right)}\right]\Gamma\left(1-\frac{2}{K}\right). \tag{S14}$$

Here, we have assumed that the length of the pore is much longer than the thermal length: $L \gg \ell = \pi v/T$ in the temperature range of interest. The scaling function has the limiting expressions

$$\Phi(z\to 0) = 2\pi^{\frac{1}{2}+\frac{2}{K}}\cos\left(\frac{\pi}{K}\right)\Gamma\left(\frac{1}{2}-\frac{1}{K}\right)\Gamma\left(\frac{1}{K}\right) \tag{S15}$$

$$\Phi(z\to\infty) = (2\pi)^{2/K}\sin\left(\frac{2\pi}{K}\right)\Gamma\left(1-\frac{2}{K}\right)z^{-2+\frac{2}{K}}. \tag{S16}$$

For comparison with experimental results, the phase slip rate $\Delta\dot\phi$ needs to be related to the pressure difference between the two reservoirs on opposite sides of the quantum channel. In analogy to contacts in electrical transport, the reservoirs are three-dimensional and in thermal equilibrium, such that the Gibbs-Duhem relation is applicable:

$$\Delta P/\rho_s = \hbar\Delta\dot\phi \tag{S17}$$

with $\rho_s$ the 3D superfluid number density. We now introduce a pressure dependent temperature $k_B T^* = \hbar\Delta P/\rho_s$ and velocity scale $v^* = k_B T^*/(\hbar\rho_0)$ with $\rho_0$ the 1D density where we have reinserted $\hbar$ and $k_B$ for clarity.

---


\fi

\end{document}